\documentstyle{article}

\def\be{\begin{equation}}
\def\ee{\end{equation}}
\def\bea{\begin{eqnarray}}
\def\eea{\end{eqnarray}}
\begin{document}

\pagestyle{empty}
\vskip-10pt
\hfill {\tt hep-th/0111150}

\begin{center}
\vskip 5truecm
{\LARGE \bf The quantum Hilbert space of a chiral two-form \\in $d = 5 + 1$ dimensions}\\ 
\vskip 2truecm
{\large \bf M{\aa}ns Henningson}\\
\vskip 1truecm
{\it Institute of Theoretical  Physics, Chalmers University of Technology\\
S-412 96 G\"{o}teborg, Sweden}\\
\vskip 5truemm
{\tt mans@fy.chalmers.se}
\end{center}
\vskip 2truecm
\noindent{\bf Abstract:}
We consider the quantum theory of a two-form gauge field on a space-time which is a direct product of time and a spatial manifold, taken to be a compact five-manifold with no torsion in its cohomology. We show that the Hilbert space of this non-chiral theory is a certain subspace of a tensor product of two spaces, that are naturally interpreted as the Hilbert spaces of a chiral and anti-chiral two-form theory respectively. We also study the observable operators in the non-chiral theory that correspond to the electric and magnetic field strengths, the Hamiltonian, and the exponentiated holonomy of the gauge-field around a spatial two-cycle. All these operators can be decomposed into contributions pertaining to the chiral and anti-chiral sectors of the theory.
\newpage
\pagestyle{plain}

\section{Introduction}
Giving a proper definition of the still rather mysterious interacting $(2, 0)$ superconformal quantum theories in $d = 5 + 1$ dimensions \cite{Witten95}\cite{Strominger} appears as a major challenge for the future. These theories are quite interesting in their own right. (See e.g. \cite{Aharony} or \cite{Seiberg} for a review.) They also seem to provide the right framework for gaining a better understanding of some aspects of $N = 4$ super Yang-Mills theory in $d = 3 + 1$ dimensions, to which they reduce upon compactification on a two-torus. Finally, the study of these theories could be viewed as an viable approach to the broader problem of understanding string and $M$-theory by studying them in a context in which the subtle conceptual difficulties of quantum gravity are absent.

At a generic point in its moduli space, an interacting $(2, 0)$ theory can be described as a set of $(2, 0)$ massless tensor multiplets strongly coupled to self-dual tensile strings that have acquired their tension by a mechanism somehow related to the Higgs mechanism for a spontaneously broken gauge symmetry. A possible approach to studying the interacting theory would thus be to first obtain a thorough understanding of the theory of free second-quantized tensor multiplets. One would then couple these, first to classical currents, then to currents from first-quantized tensile strings, and finally to currents from some second-quantized theory, possibly describing self-dual strongly interacting tensionless strings.

Already the first step in this programme, namely the quantization of a free tensor multiplet, is a rather subtle and interesting problem, in particular because of the chiral two-form gauge field (i.e. with self-dual three-form field strength) that is part of it. Such a field cannot be described by a covariant Lagrangian in the ordinary sense, but it is by now generally agreed on that the quantum theory is still well-defined as a subsector of the quantum theory of an ordinary non-chiral two-form (which of course has a Lagrangian description) \cite{Witten96}. Sofar, this procedure has mostly been carried through with a functional integral formalism in six-dimensional spaces with {\it Euclidean} signature. This means that the duality operator acting on three-forms squares to minus the identity and thus acts as a complex structure. Objects pertaining to the chiral and the anti-chiral subsector of the theory are then characterized by being holomorphic and anti-holomorphic respectively with respect to this complex structure \cite{Henningson-Nilsson-Salomonson}. 

This `holomorphic factorization' approach in Euclidean space is however not quite satisfactory. Indeed, chiral two-forms (and the symplectic Majorana spinors that are also part of the tensor multiplets) really only exist in space-times of Minkowski signature. It would therefore be desirable to directly construct the quantum theory of a chiral two-form on such a space-time. Formally, one could also here work with a functional integral formalism, but in practice this would have to be viewed as a Wick rotation of a Euclidean functional integral, and could thus only be carried through on a space-time $Y$ that is a direct product of time ${\bf R}$ and a spatial five-manifold $M$ endowed with some smooth Riemannian metric $g$, i.e.
\be
Y = {\bf R} \times M . \label{space-time}
\ee
A more serious objection is maybe that it is unclear to what extent the functional integral formalism can be applied to the case of the interacting $(2, 0)$ theories, where we do not even know whether a description in terms of a set of fields is possible or not. In this paper, we will therefore instead use the canonical formalism (which of course also limits us to space-times $Y$ of the same form as in (\ref{space-time})) to describe the quantum theory of a chiral two-form. This means that we should construct the Hilbert space of the theory, describe what the observables are, and define how the corresponding linear operators act on the Hilbert space. This way of describing a quantum theory should be sufficiently general to allow generalization to the interacting $(2, 0)$ theories. The present paper possibly appears as rather pedantic, and some of the results have already appeared in a different form in the literature, but we still hope that our methods will prove useful for future developments, in particular for understanding the coupling of a chiral theory to external currents.

Our approach to the problem of defining the quantum theory of a chiral two-form is again to view this theory as a subsector of the theory of a non-chiral two-form gauge field $B$. For the latter theory, it is straightforward to construct its Hilbert space ${\cal V}$ and describe the action of the observable operators on it. The observables that we will be considering in this paper correspond to the `electric' field-strength two-form $G$, the `magnetic' field-strength three-form $H$, the Hamiltonian ${\cal H}$, and finally the exponentiated holonomy or Wilson surface observable $W (\Sigma)$ of the two-form gauge-field around a spatial two-cycle $\Sigma$. These quantities are given by
\bea
G & = & \dot{B} \cr
H & = & d B ,
\eea
\be
{\cal H} = \frac{1}{8 \pi} \int_M \left(G \wedge * G + H \wedge * H \right) , \label{Hamiltonian}
\ee
and
\be
W (\Sigma) = \exp i \int_\Sigma B .
\ee
(Here a dot denotes the derivative with respect to time, $d$ is the exterior derivative on $M$, and * denotes the Hodge duality operator that derives from the metric $g$ on $M$ and maps two-forms and three-forms on $M$ into each other. We work in the gauge $e \cdot B = 0$, where $e$ is the vector field that generates time translations.) The quantities $G$, $H$, and ${\cal H}$ are real and correspond to Hermitian operators, whereas $W (\Sigma)$ is $U(1)$-valued and corresponds to a unitary operator.

Since the non-chiral two-form theory is free, one would expect its Hilbert space ${\cal V}$ to be the tensor product of the Hilbert spaces ${\cal V}_+$ and ${\cal V}_-$ of the chiral and anti-chiral theories. It will turn out, however, that because of a subtle correlation between these sectors, ${\cal V}$ is rather a subspace of ${\cal V}_+ \otimes {\cal V}_-$, i.e.
\be
{\cal V} \subset {\cal V}_+ \otimes {\cal V}_- .
\ee
One would also expect that the observable operators $G$, $H$, ${\cal H}$ and $W (\Sigma)$ can be decomposed into operators acting within ${\cal V}_+$ or ${\cal V}_-$ as
\bea
G & = & G_+ + G_- \cr
H & = & H_+ + H_- ,
\eea
\be
{\cal H} = {\cal H}_+ + {\cal H}_- ,
\ee
and
\be
W (\Sigma) = W_+ (\Sigma) W_- (\Sigma) .
\ee
As we will see, this is indeed true, provided that we regard $W_+ (\Sigma)$ and $W_- (\Sigma)$ as elements of the sets of unitary operators on ${\cal V}_+$ and ${\cal V}_-$ modulo $\pm 1$. (Their product $W (\Sigma)$ is well-defined as a unitary operator on ${\cal V}$, though.)

We end this introduction by discussing the choice of spatial five-manifold $M$ that we will work on. As usual in quantum mechanics, it is convenient to work on a compact manifold $M$, so that various operators have discrete rather than continuous spectra with normalizable eigenvectors that form a complete set. The simplest possibility is then of course a topologically trivial compact $M$. However, if one later wishes to couple the theory to prescribed magnetic currents supported on some one-dimensional strings, one will effectively remove the locus of these strings from $M$ and thus create a (non-compact) space with homologically non-trivial three-cycles. It is therefore natural to include the effects of non-trivial spatial topology already now. In this paper, we will consider a compact manifold $M$ with non-trivial second and third homology groups, which for simplicity we take to be torsion-free, i.e.
\be
H_2 (M, {\bf Z}) \simeq H_3 (M, {\bf Z}) \simeq {\bf Z}^b
\ee
for some integer $b$. We endow $M$ with a smooth Riemannian metric $g$.

\section{The phase space}
To quantize the theory of a non-chiral two-form gauge field $B$, it is important to note that $B$ is only locally (i.e. over a single coordinate patch) a two-form. Globally, it is rather a connection on a 1-gerbe over the space-time manifold (\ref{space-time}), and may thus undergo a gauge transformation $B \rightarrow B + \Delta B$ from one coordinate patch to another. The gauge parameter $\Delta B$ is here a closed two-form such that the de~Rham cohomology class $\left[\frac{\Delta B}{2 \pi} \right]$ is the image of an integer class on the overlap of the two coordinate patches. 

We may now proceed with a standard canonical analysis starting from a covariant Lagrangian of Maxwell type. We choose the gauge $e \cdot B = 0$, where $e$ is the vector field that generates time translations. The remaining dynamical variable is then the restriction of $B$ to $M$ at some time $t$, which we henceforth denote just $B$. The canonical momentum conjugate to $B$ is some multiple of the two-form
\be
G = \dot{B} ,
\ee
i.e. the electric field strength. This means that the equal time Poisson bracket $\left\{ . , . \right\}$ between functionals of $B$ only or functionals of $G$ only vanish, whereas
\be
\left\{ \int_\Sigma B, \int_M G \wedge * \alpha \right\} = 4 \pi \int_\Sigma \alpha \label{PB}
\ee
for an arbitrary two-cycle $\Sigma$ and an arbitrary two-form $\alpha$ on $M$. In terms of 
\be
H = d B , 
\ee
i.e. the magnetic field strength, we have that the Poisson bracket between functionals of $B$ only or functionals of $H$ only vanish, whereas
\be
\left\{ \int_M H \wedge * \beta , \int_M G \wedge * \alpha \right\} = 4 \pi \int_M \beta \wedge * d \alpha
\ee
for an arbitrary two-form $\alpha$ and an arbitrary three-form $\beta$ on $M$. It is important for the sequel that we choose the coupling constant in the Lagrangian so that the numerical constant in the right hand side of these expressions takes precisely the value $4 \pi$. The Hamiltonian ${\cal H}$ is then given by (\ref{Hamiltonian}).

The phase-space of this system is the space of connections $B$, modulo gauge transformations, that satisfy the generalized Maxwell equations
\bea
\dot{G} & = & - d^* H \cr
\dot{H} & = & d G \cr
d^* G & = & 0 \cr
d H & = &  0 . \label{Maxwell}
\eea
One must remember, though, that two gauge inequivalent connections may differ by a `flat' connection, in which case they have the same electric and magnetic fields strengths. To distinguish them, we will, in addition to the field strengths $G$ and $H$, specify also the holonomy 
\be
\int_{[\Sigma]^0} B
\ee
of $B$ around a particular chosen representive $[\Sigma]^0$ of each homology class $[\Sigma] \in H_2 (M, {\bf Z})$. Because of the gauge ambiguity of $B$, this holonomy is well-defined only as an element of ${\bf R} / 2 \pi {\bf Z}$, i.e. $\exp i \int_{\Sigma_0} B$ is a well-defined element of $U (1)$. The Wilson surface observable $W (\Sigma)$ of an arbitrary two-cycle 
\be
\Sigma = [\Sigma]^0 + \partial D \label{Sigma}
\ee
in the homology class $[\Sigma] \in H_2 (M, {\bf Z})$ can now be written in terms of these phase space variables as
\be
W (\Sigma) = \exp i \int_\Sigma B = \exp i \int_{[\Sigma]^0} B \exp i \int_D H ,
\ee
where we have used Stokes' theorem. Note that three-chain $D$ is only defined modulo a three-cycle by (\ref{Sigma}), but the periods of the magnetic field strength $H$ over such a three-cycle is a multiple of $2 \pi$, so $W (\Sigma)$ is still well-defined as a $U (1)$-valued function of $\Sigma$.

By the Hodge theorem, the last two equations in (\ref{Maxwell}) imply that we may write $G$ and $H$ as
\bea
G & = & G^0 + G^\prime \cr 
H & = & H^0 + H^\prime , \label{Hodge}
\eea
where $G_0$ and $H_0$ are harmonic two- and three-forms respectively, $G^\prime$ is a coexact two-form, and $H^\prime$ is an exact three-form. The Hamiltonian decomposes accordingly as 
\be
{\cal H} = {\cal H}^0 + {\cal H}^\prime
\ee
with
\bea
{\cal H}^0 & = & \frac{1}{8 \pi} \int_M \left( G^0 \wedge * G^0 + H^0 \wedge * H^0 \right) \cr
{\cal H}^\prime & = & \frac{1}{8 \pi} \int_M \left( G^\prime \wedge * G^\prime + H^\prime \wedge * H^\prime \right) .
\eea
Similarly, the Wilson surface observable decomposes as 
\be
W (\Sigma) = W^0 (\Sigma) W^\prime (\Sigma)
\ee
with
\bea
W^0 (\Sigma) & = & \exp i \int_{[\Sigma]^0} B \exp i \int_D H^0 \cr
W^\prime (\Sigma) & = & \exp i \int_D H^\prime .
\eea
In fact, the Hilbert space ${\cal V}$ is a tensor product of Hilbert spaces ${\cal V}^0$ and ${\cal V}^\prime$ corresponding to the harmonic and non-harmonic modes respectively:
\be
{\cal V} = {\cal V}^0 \otimes {\cal V}^\prime .
\ee
The decomposition into chiral and anti-chiral sectors described in the introduction can be carried through separately for the harmonic and non-harmonic modes.

\section{The harmonic modes}
In this section, we will construct the quantum Hilbert space ${\cal V}^0$ corresponding to the classical phase-space variables $H^0$, $G^0$ and $\exp i \int_{[\Sigma]^0} B$,where again $H^0$ and $G^0$ are the harmonic parts of the magnetic and electric field strengths, and $\exp i \int_{[\Sigma]^0} B$ is the exponentiated holonomy around a chosen representative $[\Sigma]^0$ of each homology class $[\Sigma] \in H_2 (M, {\bf Z})$. 

We have already mentioned the fact that the periods of $H$ are multiples of $2 \pi$, already at the classical level. Since the periods of the exact three-form $H^\prime$ are trivial, this means that 
\be
H^0 = 2 \pi m^0 , \label{H0}
\ee
where $m^0$ denotes the unique harmonic representative of the de~Rham cohomology class that is the image of a class $m \in H^3 (M, {\bf Z})$. In the quantum theory, this equation is interpreted as a statement about the spectrum of the corresponding operator-valued three-form.

Classically, $G^0$ can be an arbitrary harmonic two-form, and $* G^0$ is thus an arbitrary harmonic three-form. In the quantum theory, the eigenvalues of the corresponding operator-valued three-form are of the form 
\be
* G^0 = 4 \pi n^0 , \label{G0}
\ee 
where $n^0$ denotes the unique harmonic representative of the de~Rham cohomology class that is the image of a class $n \in H^3 (M, {\bf Z})$. To see this, we take the two-cycle $\Sigma$ in (\ref{PB}) to equal the chosen representative $[\Sigma]^0$ for some $[\Sigma] \in H_2 (M, {\bf Z})$, and take $\alpha$ to equal $\frac{1}{4 \pi}$ times the harmonic two-form that is dual to $[\Sigma]^0$. We then get that 
\be
\left\{ \int_{[\Sigma]^0} B, \int_M * G^0 \wedge \alpha \right\} = 1 ,
\ee
i.e. $\int_M * G^0 \wedge \alpha$ is the canonically conjugate momentum of $\int_{[\Sigma]^0} B$. But the latter phase-space variable is periodic with period $2 \pi$, so as usual in quantum theory the spectrum of $\int_M * G_0 \wedge \alpha$ is quantized in integer units, from which statement (\ref{G0}) follows.

The operators $H^0$ and $* G^0$ commute, so as an orthonormal basis in the Hilbert space ${\cal V}^0$ we may choose their simultaneous eigenvectors 
\be
\left| m, n \right> ,
\ee
where $m, n \in H^3 (M, {\bf Z})$. The corresponding eigenvalues are given by (\ref{H0}) and (\ref{G0}) respectively. These states are also eigenstates of the Hamiltonian with eigenvalues
\be
{\cal H}^0 = 2 \pi \int_M \left( \frac{1}{4} m^0 \wedge * m^0 + n^0 \wedge * n^0 \right) .
\ee
Finally we should describe the action of the Wilson surface observable operator $W^0 (\Sigma)$ on these states. From the above, it follows that
\be
W^0 (\Sigma) \left| m, n \right> = \left| m, n + 1 \right> \exp 2 \pi i \int_D m^0 .
\ee
Again, although the three-chain $D$ is only well-defined modulo a three-cycle by (\ref{Sigma}), the integral of $m^0$ over $D$ is well-defined modulo an integer, so $W^0 (\Sigma)$ is a well-defined unitary operator that only depends on $\Sigma$.

We will now describe how the Hilbert space ${\cal V}^0$ can be viewed as a subspace of ${\cal V}^0_+ \otimes {\cal V}^0_-$, where ${\cal V}^0_+$ and ${\cal V}^0_-$ are Hilbert spaces pertaining to the chiral and anti-chiral theories respectively. We first note that, by our assumption about $M$, $H^3 (M, {\bf Z}_2)$ is simply the mod~2 reduction of $H^3 (M, {\bf Z})$. We let $[k] \in H^3 (M, {\bf Z}_2)$ denote the mod~2 reduction of an element  $k \in H^3 (M, {\bf Z})$. We also choose some {\it fixed} lifting $\overline{a} \in H^3 (M, {\bf Z})$ of elements $a \in H^3 (M, {\bf Z}_2)$. (This means that $[\overline{a}] = a$ for all $a \in H^3 (M, {\bf Z}_2)$.) The spaces ${\cal V}^0_+$ and  ${\cal V}^0_-$ are spanned by the orthonormal states
\be
\left| k_+, a_+ \right> \label{V+}
\ee
and
\be
\left| k_-, a_- \right> 
\ee
respectively, where $k_+, k_- \in H^3 (M, {\bf Z})$ and $a_+ , a_- \in H^3 (M, {\bf Z}_2)$. The space ${\cal V}^0$ is now isomorphic to the subspace $\hat{\cal V} \subset {\cal V}^0_+ \otimes {\cal V}^0_-$ given by states of the form
\be
\left| k_+, a_+ \right> \otimes \left| k_-, a_- \right>
\ee
subject to the conditions
\bea
[k_+] & = & [k_-] \cr
a_+ & = & a_- .
\eea  
The isomorphism between ${\cal V}$ and $\hat{\cal V}$ is given by
\be
\left| m , n \right> = \left| k_+ , a_+ \right> \otimes \left| k_- , a_- \right>
\ee
where
\bea
m & = & k_+ + k_- + \overline{a_+} \cr
n & = & \frac{1}{2} (k_+ - k_-)
\eea
or equivalently
\bea
k_+ & = & n + \frac{1}{2} (m - \overline{[m]}) \cr
k_- & = & -n + \frac{1}{2} (m - \overline{[m]}) \cr
a_+ & = & [m] \cr
a_- & = & [m] .
\eea

This construction is justified by showing how the various observable operators can be decomposed into contributions acting within ${\cal V}_+$ or ${\cal V}_-$. Beginning with the electric and magnetic field strengths, we have $G^0 = G^0_+ + G^0_-$ and $H^0 = H^0_+ + H^0_-$. The states $\left| k_+ , a_+ \right>$ are eigenstates of $* G^0_+$ and $H^0_+$ with eigenvalues
\bea
* G^0_+ & = & \pi (2 k_+ + \overline{a_+})^0 \cr
H^0_+ & = & \pi (2 k_+ + \overline{a_+})^0 ,
\eea
so $* G^0_+ = H^0_+$ on the Hilbert space ${\cal V}^0_+$ of the chiral theory. (Superscript ${}^0$ on a class in $H^3 (M, {\bf Z})$ denotes the unique harmonic representative of the corresponding de Rham cohomology class.) Similarly, the states $\left| k_- , a_- \right>$ are eigenstates of $* G^0_-$ and $H^0_-$ with eigenvalues
\bea
* G^0_- = - \pi (2 k_- + \overline{a_-})^0 \cr
H^0_- = \pi (2 k_- + \overline{a_-})^0 ,
\eea 
so that $* G^0_- = - H^0_-$ on ${\cal V}^0_-$ as is appropriate for the anti-chiral theory. One should note that the periods of $H^0_+$ and $H^0_-$ may be half-integer multiples of $2 \pi$ as in \cite{Witten99}. The states $\left| k_+ , a_+ \right>$ and $\left| k_- , a_- \right>$ are also eigenstates of the Hamiltonians of the chiral and anti-chiral theories respectively with eigenvalues
\bea
{\cal H}^0_+ & = & \frac{\pi}{4} \int_M (2 k_+ + \overline{a_+})^0 \wedge * (2 k_+ + \overline{a_+})^0 \cr
{\cal H}^0_- & = & \frac{\pi}{4} \int_M (2 k_- + \overline{a_-})^0 \wedge * (2 k_- + \overline{a_-})^0
\eea
so that indeed ${\cal H}^0 = {\cal H}^0_+ + {\cal H}^0_-$. Finally we decompose the Wilson surface observable $W^0 (\Sigma)$ as $W^0 (\Sigma) = W^0_+ (\Sigma) W^0_- (\Sigma)$ with
\bea
W^0_+ (\Sigma) \left| k_+ , a_+ \right> & = & \left| k_+ + 1 , a_+ \right> \exp i \pi \int_D (2 k_+ + \overline{a_+})^0 \cr
W^0_- (\Sigma) \left| k_- , a_- \right> & = & \left| k_- - 1 , a_- \right> \exp i \pi \int_D (2 k_- + \overline{a_-})^0 .
\eea
But since $D$ is only well-defined modulo a three-cycle by (\ref{Sigma}), $W^0_+ (\Sigma)$ and $W^0_- (\Sigma)$ are only well-defined as unitary operators modulo $\pm 1$. However, their product $W^0 (\Sigma)$ is of course well-defined as a unitary operator on $\hat{\cal V} \simeq {\cal V}$, since $a_+ = a_-$ on this space.

Finally, we note that ${\cal V}^0_+$ may be decomposed as
\be
{\cal V}^0_+ = \bigoplus_{a_+ \in H^3 (M, {\bf Z})} {\cal V}^0_{a_+} , 
\ee
where ${\cal V}^0_{a_+}$ is spanned by states of the form (\ref{V+}) for $k_+ \in H^3 (M , {\bf Z})$. The observables that we have been considering do not mix these subspaces, each of which could thus be interpreted as the Hilbert space of a different version of the chiral theory. The fact that there are inequivalent chiral theories was first seen on a general six-dimensional space-time in \cite{Witten96}, where it was pointed out that there is no canonical way of picking one of the theories. In our case, where the six-dimensional general covariance is explicitly broken by the form of the space-time (\ref{space-time}), there is however a distingished theory, namely the $a_+ = 0$ theory.  

\section{The non-harmonic modes}
We now turn our attention to the phase-space variables $G^\prime$ and $H^\prime$, i.e. the coexact part of the electric field strength and the exact part of the magnetic field strength respectively. Our aim is to construct the Hilbert space ${\cal V}^\prime_+$ of the chiral sector of the theory and describe the action of the various observable operators on it.
 
To begin with, we will discuss the spectra of the Laplacians
\bea
\Delta^{(2)}_{\rm coexact} & = & d d^* + d^* d = d^* d \cr
\Delta^{(3)}_{\rm exact} & = & d d^* + d^* d = d d^*
\eea
acting on coexact two-forms and exact three-forms on $M$ respectively. They can be described as follows: The eigenvalues of both Laplacians are of the form $\omega^2$ for $\omega \in \Omega$, where $\Omega$ is some discrete set of positive numbers. For each $\omega \in \Omega$ there are two linearly independent coexact two-forms $G_\omega$ and $\hat{G}_\omega$ that are eigenvectors of $\Delta^{(2)}_{\rm coexact}$ and two linearly independent exact three-forms $H_\omega$ and $\hat{H}_\omega$ that are eigenvectors of $\Delta^{(3)}_{\rm exact}$. We can choose them to fulfill the orthonormality conditions
\bea
\int_M G_\omega \wedge * G_{\omega^\prime} & = & \delta_{\omega \omega^\prime} \cr
\int_M \hat{G}_\omega \wedge * \hat{G}_{\omega^\prime} & = & \delta_{\omega \omega^\prime} \cr
\int_M G_\omega \wedge * \hat{G}_{\omega^\prime} & = & 0
\eea
and
\bea
\int_M H_\omega \wedge * H_{\omega^\prime} & = & \delta_{\omega \omega^\prime} \cr
\int_M \hat{H}_\omega \wedge * \hat{H}_{\omega^\prime} & = & \delta_{\omega \omega^\prime} \cr
\int_M H_\omega \wedge * \hat{H}_{\omega^\prime} & = & 0 .
\eea
There are also the corresponding completeness relations of $G_\omega$ and $\hat{G}_\omega$ for $\omega \in \Omega$ in the space of coexact two-forms, and $H_\omega$ and $\hat{H}_\omega$ for $\omega \in \Omega$ in the space of exact three-forms. The forms $G_\omega$, $\hat{G}_\omega$, $H_\omega$ and $\hat{H}_\omega$ are related to each other in the following way:
\be
\begin{array}{ccc}
& {\scriptstyle \frac{1}{\omega} * d} & \cr
G_\omega & \stackrel{\longleftarrow}{\longrightarrow} & \hat{G}_\omega \cr
& {\scriptstyle - \frac{1}{\omega} * d} & \cr
{\scriptstyle \frac{1}{\omega} d} \downarrow \uparrow {\scriptstyle \frac{1}{\omega} d^*} & & {\scriptstyle \frac{1}{\omega} d} \downarrow \uparrow {\scriptstyle \frac{1}{\omega} d^*} \cr
& {\scriptstyle \frac{1}{\omega} d *} & \cr
H_\omega & \stackrel{\longleftarrow}{\longrightarrow} & \hat{H}_\omega \cr
& {\scriptstyle - \frac{1}{\omega} d *} &
\end{array} .
\ee
Thus $H_\omega = * \hat{G}_\omega$ and $\hat{H}_\omega = - * G_\omega$.

We can now write the general solution to the equations (\ref{Maxwell}) in the form (\ref{Hodge}) with
\bea
G^\prime & = & \sum_{\omega \in \Omega} \left(\left(a_\omega \cos \omega t - b_\omega \sin \omega t\right) G_\omega + \left(\hat{a}_\omega \cos \omega t - \hat{b}_\omega \sin \omega t\right) \hat{G}_\omega \right) \cr
H^\prime & = & \sum_{\omega \in \Omega} \left(\left(a_\omega \sin \omega t + b_\omega \cos \omega t\right) H_\omega + \left(\hat{a}_\omega \sin \omega t + \hat{b}_\omega \cos \omega t\right) \hat{H}_\omega \right) ,
\eea
where $a_\omega$, $b_\omega$, $\hat{a}_\omega$ and $\hat{b}_\omega$ are arbitrary constants. In particular, at time $t = 0$, we have
\bea
G^\prime |_{t = 0} & = & \sum_{\omega \in \Omega} \left(a_\omega G_\omega + \hat{a}_\omega \hat{G}_\omega \right) \cr
H^\prime |_{t = 0} & = & \sum_{\omega \in \Omega} \left(b_\omega H_\omega + \hat{b}_\omega \hat{H}_\omega \right) ,
\eea
so that
\bea
a_\omega & = & \int_M G \wedge * G_\omega \cr
b_\omega & = & \int_M H \wedge * H_\omega \cr
\hat{a}_\omega & = & \int_M G \wedge * \hat{G}_\omega \cr
\hat{b}_\omega & = & \int_M H \wedge * \hat{H}_\omega 
\eea
with the integrals evaluated at $t = 0$. It then follows from (\ref{PB}) that the only non-vanishing Poisson brackets between the variables $a_\omega$, $b_\omega$, $\hat{a}_\omega$ and $\hat{b}_\omega$ are
\bea
\left\{a_\omega , b_{\omega^\prime} \right\} & = & - 4 \pi \omega \delta_{\omega , \omega^\prime} \cr
\left\{\hat{a}_\omega , \hat{b}_{\omega^\prime} \right\} & = & - 4 \pi \omega \delta_{\omega , \omega^\prime} .
\eea

In order to disentangle the chiral and anti-chiral sectors of the theory, we introduce the linear combinations $c^+_\omega$, $c^-_\omega$, $\hat{c}^+_\omega$, and $\hat{c}^-_\omega$ defined by
\bea
c^\pm_\omega & = & \frac{1}{\sqrt{2}} (a_\omega \mp \hat{b}_\omega) \cr
\hat{c}^\pm_\omega & = & \frac{1}{\sqrt{2}} (\hat{a}_\omega \pm b_\omega) .
\eea 
The only non-vanishing Poisson brackets between these variables are
\bea
\left\{ c^+_\omega, \hat{c}^+_{\omega^\prime} \right\} & = & - 4 \pi \omega \delta_{\omega , \omega^\prime} \cr
\left\{ c^-_\omega, \hat{c}^-_{\omega^\prime} \right\} & = & 4 \pi \omega \delta_{\omega , \omega^\prime} .
\eea
We can now decompose the electric and magnetic field strengths into contributions from the chiral and anti-chiral sectors of the theory, i.e. $G^\prime = G^\prime_+ + G^\prime_-$ and $H^\prime = H^\prime_+ + H^\prime_-$, where
\bea
H^\prime_+ & = & \frac{1}{\sqrt{2}} \sum_{\omega \in \Omega} \left(\left(\hat{c}^+_\omega \cos \omega t + c^+_\omega \sin \omega t \right) H_\omega + \left(- c^+_\omega \cos \omega t + \hat{c}^+_\omega \sin \omega t \right) \hat{H}_\omega \right) \cr
G^\prime_+ & = & \frac{1}{\sqrt{2}} \sum_{\omega \in \Omega} \left(\left(c^+_\omega \cos \omega t - \hat{}c^+_\omega \sin \omega t \right) G_\omega + \left(\hat{c}^+_\omega \cos \omega t + c^+_\omega \sin \omega t \right) \hat{G}_\omega \right) 
\eea
and
\bea
H^\prime_- & = & \frac{1}{\sqrt{2}} \sum_{\omega \in \Omega} \left(\left(- \hat{c}^+_\omega \cos \omega t + c^+_\omega \sin \omega t \right) H_\omega + \left(c^+_\omega \cos \omega t + \hat{c}^+_\omega \sin \omega t \right) \hat{H}_\omega \right) \cr
G^\prime_- & = & \frac{1}{\sqrt{2}} \sum_{\omega \in \Omega} \left(\left(c^+_\omega \cos \omega t + \hat{c}^+_\omega \sin \omega t \right) G_\omega + \left(\hat{c}^+_\omega \cos \omega t - c^+_\omega \sin \omega t \right) \hat{G}_\omega \right) .
\eea
We see that indeed $* G^\prime_+ = H^\prime_+$ and $* G^\prime_- = - H^\prime_-$ as desired. The Hamiltonian also decomposes as ${\cal H}^\prime = {\cal H}^\prime_+ + {\cal H}^\prime_-$, where
\bea
{\cal H}^\prime_+  & = & \frac{1}{8 \pi} \sum_{\omega \in \Omega} \left((c^+_\omega)^2 + (\hat{c}^+_\omega)^2 \right) \cr
{\cal H}^\prime_-  & = & \frac{1}{8 \pi} \sum_{\omega \in \Omega} \left((c^-_\omega)^2 + (\hat{c}^-_\omega)^2 \right) .
\eea
Finally, we decompose the Wilson surface observable as $W^\prime (\Sigma) = W^\prime_+ (\Sigma) W^\prime_- (\Sigma)$, where
\bea
W^\prime_+ (\Sigma) & = & \exp i \int_D H^\prime_+ \cr
& = & \exp \frac{i}{\sqrt{2}} \sum_{\omega \in \Omega} \frac{1}{\omega} \int_{\Sigma - [\Sigma]^0} \left(\left( \hat{c}^+_\omega \cos \omega t + c^+_\omega \sin \omega t \right) G_\omega + \left(- c^+_\omega \cos \omega t + \hat{c}^+_\omega \sin \omega t \right) \hat{G}_\omega \right) \cr
W^\prime_- (\Sigma) & = & \exp i \int_D H^\prime_- \cr
& = & \exp \frac{i}{\sqrt{2}} \sum_{\omega \in \Omega} \frac{1}{\omega} \int_{\Sigma - [\Sigma]^0} \left(\left(- \hat{c}^-_\omega \cos \omega t + c^-_\omega \sin \omega t \right) G_\omega + \left(c^-_\omega \cos \omega t + \hat{c}^-_\omega \sin \omega t \right) \hat{G}_\omega \right) .
\eea
Here we have used the exactness of $H_\omega$ and $\hat{H}_\omega$ to write $W^\prime_+ (\Sigma)$ and $W^\prime_- (\Sigma)$ in a form where it is manifest that the choice of $D$ fulfilling (\ref{Sigma}) is immaterial.

By the usual procedure of canonical quantization, real functions on the classical phase space correspond to Hermitian operators on the quantum Hilbert space, with the operator commutator $\left[ . , . \right]$ related to the Poisson bracket as $i \left\{ . , . \right\}$. We will now carry out this procedure for the chiral sector of the theory. We introduce the annihilitation operators $\alpha_\omega$ and their Hermitian conjugates the creation operators $\alpha_\omega^\dagger$ for $\omega \in \Omega$ as
\bea
\alpha_\omega & = & \frac{1}{\sqrt{8 \pi \omega}} (c^+_\omega + i \hat{c}^+_\omega) \cr
\alpha_\omega^\dagger & = & \frac{1}{\sqrt{8 \pi \omega}} (c^+_\omega - i \hat{c}^+_\omega) .
\eea
They obey the harmonic oscillator commutation relations
\bea
[\alpha_\omega , \alpha_{\omega^\prime}] & = & 0 \cr
[\alpha_\omega^\dagger , \alpha_{\omega^\prime}^\dagger] & = & 0 \cr
[\alpha_\omega , \alpha_{\omega^\prime}^\dagger] & = & \delta_{\omega, \omega^\prime} .
\eea
This algebra can be represented on a Hilbert space with basis vectors of the form
\be
\left| \left\{ n_\omega \right\} \right> = \bigotimes_{\omega \in \Omega} \left| n_\omega \right> = \bigotimes_{\omega \in \Omega} \left(\alpha^\dagger_\omega \right)^{n_\omega} \left| 0 \right> ,
\ee
where the $n_\omega$ are non-negative integers. These states are eigenvectors of the normal-ordered occupation-number operators $N_\omega = \alpha^\dagger_\omega \alpha_\omega$ with eigenvalues $n_\omega$.

We now turn to the construction of observable operators that act on this Hilbert space. The magnetic and electric field strength operators are straightforward:
\bea
H^\prime_+ & = & \sum_{\omega \in \Omega} \sqrt{\pi \omega} \left(\alpha^\dagger_\omega e^{- i \omega t} (- \hat{H}_\omega + i H_\omega) + \alpha_\omega e^{i \omega t} (- \hat{H}_\omega - i H_\omega) \right) \cr
G^\prime_+ & = & \sum_{\omega \in \Omega} \sqrt{\pi \omega} \left(\alpha^\dagger_\omega e^{- i \omega t} (G_\omega + i \hat{G}_\omega) + \alpha_\omega e^{i \omega t} (G_\omega - i \hat{G}_\omega) \right) . \label{GHoperators}
\eea
Being bilinear in the field strengths, the Hamilton operator suffers from an ordering ambiguity. We fix this by the usual choice of zero-point energy so that
\be
{\cal H}^\prime_+ = \sum_{\omega \in \Omega} \omega \left(N_\omega + \frac{1}{2} \right) .
\ee
The exponentiated holonomy operator of course suffers from a more serious ordering problem. In this paper, we will define it as
\be
W^\prime_+ (\Sigma) = \exp i \int_D H^\prime_+
\ee
with $H^\prime_+$ given by (\ref{GHoperators}). Again, we may of course use the exactness of $H^\prime_+$ to write $W^\prime_+ (\Sigma)$ in form where the apparent dependence on a choice of $D$ disappears. 

By construction, $W^\prime_+ (\Sigma)$ is formally unitary, i.e. 
\be
W^\prime_+ (\Sigma) (W^\prime_+ (\Sigma))^\dagger = (W^\prime_+ (\Sigma))^\dagger W^\prime_+ (\Sigma) = 1 .
\ee
If we introduce a second two-cycle $\Sigma^\prime = [\Sigma^\prime]^0 + \partial D^\prime$, the operators $W^\prime_+ (\Sigma)$, $W^\prime_+ (\Sigma^\prime)$, and $W^\prime_+ (\Sigma + \Sigma^\prime)$ fulfill an interesting algebra. To investigate this, we start by computing the commutator of the operators $\int_D H^\prime_+$ and $\int_{D^\prime} H^\prime_+$:
\bea
\left[ \int_D H^\prime_+ \right. , && \left. \!\!\!\!\!\!\!\!\!\!\!\!\!\!\! \int_{D^\prime} H^\prime_+ \right] \cr
& = & 2 \pi i \sum_{\omega \in \Omega} \omega \left( \int_D H_\omega \int_{D^\prime} \hat{H}_\omega - \int_D \hat{H}_\omega \int_{D^\prime} H_\omega \right) \cr
& = & 2 \pi i \sum_{\omega \in \Omega} \left( \int_D H_\omega \int_{D^\prime} d * H_\omega + \int_D \hat{H}_\omega \int_{D^\prime} d * \hat{H}_\omega \right) \cr
& = & 2 \pi i \sum_{\omega \in \Omega} \left( \int_D H_\omega \int_{\Sigma^\prime - [\Sigma^\prime]^0} * H_\omega + \int_D \hat{H}_\omega \int_{\Sigma^\prime - [\Sigma^\prime]^0} * \hat{H}_\omega \right) \cr
& = & 2 \pi i \sum_{\omega \in \Omega} \left( \int_M * P_D \wedge * H_\omega \int_M P_{\Sigma^\prime - [\Sigma^\prime]^0} \wedge * H_\omega + \int_M * P_D \wedge * \hat{H}_\omega \int_M P_{\Sigma^\prime - [\Sigma^\prime]^0} \wedge * \hat{H}_\omega \right) \cr
& = & 2 \pi i \int_M P_D \wedge P_{\Sigma^\prime - [\Sigma^\prime]^0} \cr
& = & 2 \pi i D \cdot (\Sigma^\prime - [\Sigma^\prime]^0) \cr
& = & 2 \pi i L (\Sigma - [\Sigma]^0 , \Sigma^\prime - [\Sigma^\prime]^0) .
\eea
In the fourth line, we have introduced the Poincar\'e duals $P_D$ and $P_{\Sigma^\prime - [\Sigma^\prime]^0}$ of $D$ and $\Sigma^\prime - [\Sigma^\prime]^0$ respectively, and in the next line we have used the completeness relation of $H_\omega$ and $\hat{H}_\omega$ for $\omega \in \Omega$ in the space of exact three-forms on $M$. In the sixth line, $D \cdot (\Sigma^\prime - [\Sigma^\prime]^0)$ denotes the intersection number of $D$ and $\Sigma^\prime - [\Sigma^\prime]^0$, which by definition equals the linking number $L(\Sigma - \Sigma^0 , \Sigma^\prime - [\Sigma^\prime]^0)$ of $\Sigma - [\Sigma]^0$ and $\Sigma^\prime - [\Sigma^\prime]^0$. This is an integer, so the commutator we have calculated is a multiple of $2 \pi i$. (Essentially the same calculation was performed in \cite{Henningson} for the case when $M$ is flat Euclidean five-space.) It now follows from the Baker-Hausdorff formula that 
\be
W^\prime_+ (\Sigma) W^\prime_+ (\Sigma^\prime) = W^\prime_+ (\Sigma^\prime) W^\prime_+ (\Sigma) = (-1)^{L (\Sigma - [\Sigma]^0, \Sigma^\prime - [\Sigma^\prime]^0)} W^\prime_+ (\Sigma + \Sigma^\prime) , \label{multiplication}
\ee
i.e. $W^\prime_+ (\Sigma)$ and $W^\prime_+ (\Sigma^\prime)$ commute, and their product is up to a sign equal to $W^\prime_+ (\Sigma + \Sigma^\prime)$. One should note that this sign in the multiplication law appears also on a manifold $M$ of trivial topology, in contrast to the sign ambiguity in the {\it definition} of $W^0_+ (\Sigma)$ on a topologically non-trivial $M$ that we discussed in the previous section. Altogether, on a topologically non-trivial manifold, we should think of the complete operator $W_+ (\Sigma) = W^0_+ (\Sigma) W^\prime_+ (\Sigma)$ as an element of the set of unitary operators modulo $\pm 1$, in which case the the sign factor in the multiplication law (\ref{multiplication}) of course is irrelevant.

Finally, we will briefly discuss the divergences of the operator $W^\prime_+ (\Sigma)$. This operator is trivially a product over $\omega \in \Omega$ of commuting factors $W^\prime_\omega (\Sigma)$. Each of these may by the Baker-Hausdorff formula be rewritten as a product of an {\it Hermitian} $c$-number pre-factor, a factor involving the creation operator $\alpha^\dagger_\omega$, and a factor involving the annihilation operator $\alpha_\omega$ (in that order). It is then clear that $W^\prime_\omega (\Sigma)$ has finite matrix elements $\left< n_\omega \right| W_\omega^\prime (\Sigma) \left| n^\prime_\omega \right>$ (which for the case of $n_\omega = n^\prime_\omega = 0$ equals the $c$-number prefactor). However, the infinite product over $\omega \in \Omega$ will in general diverge. To get a finite result, one would have to regulate the product (e.g. by introducing some kind of cutoff on $\omega$) and then modify the definition of $W^\prime (\Sigma)$ by including some Hermitian $c$-number factor so that the regulator may be removed. Some of the formal properties of $W^\prime (\Sigma)$, such as its unitarity and also its scale invariance, will be lost in this process, though. (See \cite{Henningson-Skenderis} for a related discussion on the Weyl anomaly of Wilson surface observables in the non-chiral theory.)

\vspace*{5mm}
This research was supported by the Swedish Research Council.

\end{document}